# Dynamical stability analysis of tumor growth and invasion: A reaction-diffusion model

Ahmed M. Fouad

*Department of Physics, Temple University, Philadelphia, PA 19122, U.S.A.*

The acid-mediated tumor invasion hypothesis proposes that altered glucose metabolism exhibited by the vast majority of tumors leads to increased acid ($H^+$ ion) production which subsequently facilitates tumor invasion [1-3]. The reaction-diffusion model [2] that captures the key elements of the hypothesis shows how the densities of normal cells, tumor cells, and excess $H^+$ ions change with time due to both chemical reactions between these three populations and density-dependent diffusion by which they spread out in three-dimensional space. Moreover, it proposes that each cell has an optimal pH for survival; that is, if the local pH deviates from the optimal value in either an acidic or alkaline direction, the cells begin to die, and that the death rate saturates at some maximum value when the microenvironment is extremely acidic or alkaline. We have previously studied in detail how the death-rate functions of the normal and tumor populations depend upon the $H^+$ ion density [4]. Here, we extend previous work by investigating how the equilibrium densities (at which the time rates of change of the cellular densities are equal to zero) reached by the normal and tumor populations in three-dimensional space are affected by the presence of the $H^+$ ions, and we present detailed analytical and computational techniques to analyze the dynamical stability of these equilibrium densities. For a sample set of biological input parameters and within the acid-mediation hypothesis, our model predicts the transformation to a malignant behavior, as indicated by the presence of unstable sets of equilibrium densities.

**Keywords:** Dynamical stability analysis, Fixed points, Cancer biology, Acid-mediated tumor invasion, Reaction-diffusion systems, Partial differential equations

## 1. Introduction

Since the pioneering work of Warburg in the 1920's [5], it has been well established that tumors consistently rely on anaerobic pathways to convert glucose to ATP, even in the presence of abundant oxygen. Because anaerobic metabolism of glucose to lactic acid is substantially less efficient than oxidation to $CO_2$ and $H_2O$, tumor cells maintain ATP production by increasing the glucose flux. The latter forms the basis for tumor imaging with FDG-PET [6-10]. This technique is now being widely applied to human cancers and has confirmed that the vast majority of primary and metastatic tumors demonstrate substantially increased glucose uptake compared to normal tissue. Furthermore, PET imaging has also demonstrated a direct correlation between tumor aggressiveness (and prognosis) and the rate of glucose consumption [6,7]. A critical consequence of this altered metabolism is increased tumor cell acid production. Initially, it was assumed that the lactic acid produced by anaerobic metabolism would result in an acidic intracellular $pH$; however, studies with NMR spectroscopy have now consistently demonstrated that the intracellular $pH$ of tumors is generally slightly alkaline compared to normal cells [11]. Other measurements have demonstrated that tumor cells actually excrete protons through amplification of the $Na^+/H^+$ antiporter and other membrane transporters. As a result, it is now clear that the extracellular $pH$ of tumors is substantially lower (usually by about 0.5 $pH$ unit) than normal [12-15]. Generally, a persistent $pH$ below about 7.1 results in death of normal cells due to a p53-dependent apoptosis pathway triggered by increased caspase activity [16,17]. However, tumor cells are relatively resistant to acidic $pH$, presumably due to mutations in p53 or other components of the apoptotic pathway. In fact, tumor cells typically exhibit a maximum proliferation rate in relatively acidic mediums (i.e. $pH$ 6.8) [18-20]. As a consequence, it has been hypothesized that tumor cells become invasive because they perturb the environment so that it is optimal for their proliferation and toxic to normal cells with which they compete for space and substrate.

An acid-mediated tumor invasion hypothesis (AMTIH) has been developed [1-3], wherein microenvironmental changes at the tumor-host interface caused by altered metabolism in transformed cells (accelerated uptake of glucose due to increased reliance on glycolysis) lead to an excess acid production, thus facilitating tumor invasion. Concurrent with this increase in glucose consumption is the up-regulation of the $Na^+/H^+$ antiport system [21,22]. The resulting net increase in tumor intracellular $pH$ leads to normalization, and therefore to a decrease in extracellular $pH$ by about 0.5 $pH$ unit compared to normal tissue [11-13,15,20]. While some investigators have attributed this increased reliance on glycolysis to the existence of a compromised tumor blood supply leading to poor $O_2$ perfusion of tumor tissue [23], others have found evidence that it is a phenotypical consequence of transformation [24].

The reaction-diffusion model of the AMTIH [2] predicts the degree to which excess hydrogen ions in the tumor interstitium diffuse into the surrounding tissue. It also predicts the extent of the zone of altered interstitial $pH$ found at the tumor-host interface. This microenvironment is harmful to normal tissue because of three reasons. First, normal cell viability declines sharply below extracellular $pH$ of 7.1 [11-13,15,20]. Second, an acidic microenvironment stimulates production of enzymes which degrade the extracellular matrix [25,26]. Third, diminished $pH$ results in loss of intercellular gap junctions reducing the cohesion, collaboration, and communication among normal cells at the tumor edge [27]. This altered microenvironment leads to a progressive loss of normal cells and degradation of extracellular matrix. This subsequently allows tumor cells (which are much more resistant to acidic extracellular $pH$) to invade the host tissue. Eventually, the acid-mediation hypothesis of tumor invasion is supported by recent reports that high lactate levels are directly correlated with the likelihood of metastases, tumor recurrence, and restricted patient survival [28].

The main research focus in this paper is to investigate how the equilibrium densities reached by the normal and tumor populations are affected by the existence of the $H^+$ ions. To achieve that, given some sample values of biological input parameters (which are patient-specific), we present a detailed approach (that utilizes both analytical and computational techniques) to solving numerically for the equilibrium points (fixed points) that represent the equilibrium densities at which the normal, tumor, and acid populations could coexist in three-dimensional space. Moreover, through solving numerically for the eigenvalues of the Jacobian matrix of the partial differential equations that represent how the three populations evolve with time, we determine the dynamical stability of these fixed points.

This article is outlined as follows. In Sec. 2 we review the previous work on modeling tumor growth and invasion. The canonical Gatenby-Gawlinski reaction-diffusion model that captures the key components of the AMTIH is presented in Sec. 3. In Sec. 4 we present our analytical and computational results on the dynamical stability analysis of the reaction-diffusion model. The conclusions and outlook are presented in Sec. 5.

## 2. Previous work

Tumor growth and other biological systems are particularly amenable to modeling with cellular automata. This approach has been used increasingly since Wolfram (1986) demonstrated that every automaton with local interactions belongs to one of a small number of universality classes [29]. Some of the earliest work in tumor modeling with cellular automata was carried out by Düchting and Vogelsaenger to investigate the effects of different radio-therapeutic strategies on tumor growth [30,31].

More recently, Qi *et al.* [32] developed a cellular automaton model of tumor growth which includes immune system surveillance and accounts for mechanical pressure from within the tumor. This model is of interest because it reproduces the Gompertz growth law. The cellular automaton approach was also employed by Smolle and Stettner [33], and by Smolle [34,35] in an attempt to infer relationships between tumor morphology and basic biological features; such as cell cohesion, motility, and autocrine and paracrine growth factors. Kansal *et al.* [36,37] have made use of the Voronoi tessellation of space in a cellular automaton model of brain tumor growth dynamics. This elegant approach allows for the preservation of the discrete nature of cells (or groups of cells); but removes the anisotropy of a regular lattice.

Cellular automata have also been used to model neoangiogenesis, which is another crucial aspect of cancer biology. Chaplain and Anderson [38,39] have used cellular automata to solve the discrete version of a system of non-linear partial differential equations which describe the response in space and time of endothelial cells to tumor angiogenesis factor and fibronectin, including migration, proliferation, anastomosis, and branching. Cellular automata have also been used to model the development of leaf veins, insect trachea, and tumor neovascularization [40].

Some models that use systems of partial differential equations have been developed to simulate tumor growth and invasion; such as tumor angiogenesis [38,39,42,46], post-surgical response to tumor removal [41], avascular tumor growth [43,44], tumor cell motility during invasion [45], and drug resistance and vascular structure in chemotherapy [47]. All these models (either automaton-based or continuum, stochastic or deterministic) simulate one or more aspects of cancer biology. Other models, including chemotherapeutic agent transfer [47] and diffusive vasculature for nutrient transfer [48], do so with a spatially-uniform term accounting for delivery or removal, not by including individual vessels distributed throughout space.

## 3. Reaction-diffusion model of tumor invasion

Gatenby and Gawlinski hypothesize that tumor-induced alteration of microenvironmental $pH$ may provide a simple, yet complete mechanism for cancer invasion. In this section, we extend a reaction-diffusion model they developed, which describes the spatial distribution and temporal development of tumor tissue, normal tissue, and excess $H^+$ ion density [2]. The model predicts a $pH$ gradient extending from the tumor-host interface, which is confirmed by reanalysis of existing experimental data. Investigation of the structure and dynamics of the tumor-host interaction within the context of the model demonstrates a transition from benign to malignant growth, analogous to the adenoma-carcinoma sequence. The effects of biological parameters critical to controlling this transition are supported by experimental and clinical observations. It is important to point out that this model describes the interaction between a growing tumor and surrounding normal tissue only in the immediate vicinity of the tumor-host interface.

Any broadly applicable model should incorporate as key parameters properties common to all or nearly all tumors, despite their underlying genetic instability and heterogeneity. One consistent cellular dynamic is the evolution of tumor populations away from the differentiated state of the tissue of origin toward one that is more primitive and less ordered. Metabolic changes parallel this evolution with increasing use of glycolytic metabolism (even in the presence of oxygen) despite the 19-fold decrease in energy production that results [2]. The Gatenby-Gawlinski model assumes that it is precisely this inefficiency which may facilitate successful tumor invasion, specifically the transformation-induced reversion of neoplastic tissue to primitive glycolytic metabolic pathways with a resultant increased acid production and the diffusion of that acid into the surrounding healthy tissue. This, in turn, creates a peritumoral microenvironment in which tumor cells survive and proliferate, whereas normal cells are unable to remain viable in such a microenvironment. This leads to the following temporal sequence. First, high $H^+$ ion concentrations in tumors extend (by chemical diffusion) into adjacent normal tissue exposing these normal cells to tumor-like interstitial $pH$. Second, normal cells immediately adjacent to the tumor edge are unable to survive in this chronically acidic environment. Lastly, the progressive loss of layers of normal cells at the tumor-host interface facilitates the tumor invasion process.

From the temporal sequence mentioned above, we can conclude that the essential elements of this tumor invasion mechanism are low interstitial $pH$ of tumors due to primitive metabolism and reduced viability of normal tissue in a $pH$ environment favorable to tumor tissue. The Gatenby-Gawlinski model mathematically frames the acid-mediation hypothesis as a system of reaction-diffusion equations which, when solved, make detailed predictions of the structure and dynamics of the tumor-host interface.

The acid-mediated tumor invasion hypothesis described above can be represented by the following system of reaction-diffusion equations. Let $N_1$ and $N_2$ denote the cell densities in $cell/cm^3$ of the normal and tumor populations, respectively. Assuming these populations only compete for available space, their temporal evolution is governed by the equations

$$\frac{\partial N_1}{\partial t} = r_1 N_1 (1 - \frac{N_1}{K_1} - \frac{N_2}{K_2}),$$

$$\frac{\partial N_2}{\partial t} = r_2 N_2 (1 - \frac{N_1}{K_1} - \frac{N_2}{K_2}), \tag{1}$$

where $r_{1,2}$ and $K_{1,2}$ are the growth rates in $s^{-1}$ ($s$ denotes second) and spatial carrying capacities in $cell/cm^3$ of the respective populations. Moreover, if we assume that cells can migrate through space via a process akin to Fickian diffusion, where the diffusion parameters are themselves density-dependent (having a maximum value in empty space and going to zero when cells are close-packed), then Eq. (1) becomes

$$\frac{\partial N_1}{\partial t} = r_1 N_1 (1 - \frac{N_1}{K_1} - \frac{N_2}{K_2}) + \nabla \cdot [D_{N_1}(1 - \frac{N_1}{K_1} - \frac{N_2}{K_2})\nabla N_1],$$

$$\frac{\partial N_2}{\partial t} = r_2 N_2 (1 - \frac{N_1}{K_1} - \frac{N_2}{K_2}) + \nabla \cdot [D_{N_2}(1 - \frac{N_1}{K_1} - \frac{N_2}{K_2})\nabla N_2], \tag{2}$$

where $D_{N_1}$ and $D_{N_2}$ in $cm^2/s$ are the empty-space diffusion constants of the normal and tumor cells, respectively. For simplicity we assume that these are approximately equal: $D_{N_1} \approx D_{N_2} = D_N$. The Lotka-Volterra terms ensure that the density-dependent diffusion parameters are always positive-definite ($\in [0, D_N]$) [49].

Next, the model assumes that each cell has an optimal $pH$ for survival, and that if the local $pH$ deviates from the optimal value in either an acidic or alkaline direction, the cells begin to die. Moreover, it assumes that the death rate saturates at some maximum value when the environment is extremely acidic or alkaline. A simple $ad$-$hoc$ functional form meeting these criteria is the "inverted Gaussian"

$$f_{1,2}(H) = d_{1,2}[1 - \exp\{-(\frac{H - H_{1,2}^{opt}}{2H_{1,2}^{width}})^2\}], \tag{3}$$

where $H$ is the local concentration of $H^+$ ions in $mol/L$, $d_{1,2}$ are the saturated death rates in $s^{-1}$, $H_{1,2}^{opt}$ are the local $H^+$ ion concentrations in $mol/L$, corresponding to the optimal $pH$'s, and $H_{1,2}^{width}$ are the half-widths of the "inverted Gaussian" in $mol/L$. Substituting the death rates (3) into (2) we obtain

$$\frac{\partial N_1}{\partial t} = r_1 N_1 (1 - \frac{N_1}{K_1} - \frac{N_2}{K_2}) - f_1(H) N_1 + D_N \nabla \cdot [(1 - \frac{N_1}{K_1} - \frac{N_2}{K_2}) \nabla N_1],$$

$$\frac{\partial N_2}{\partial t} = r_2 N_2 (1 - \frac{N_1}{K_1} - \frac{N_2}{K_2}) - f_2(H) N_2 + D_N \nabla \cdot [(1 - \frac{N_1}{K_1} - \frac{N_2}{K_2}) \nabla N_2], \tag{4}$$

The $H^+$ ions are produced at a rate proportional to the local concentration of tumor and removed by the combined effects of buffering and vascular evacuation, both of which are proportional to micro-vessel areal density. Thus

$$\frac{\partial H}{\partial t} = r_3 N_2 - d_3 (H - H_0) + D_3 \nabla^2 H, \tag{5}$$

where $H$ is the $H^+$ ion concentration in $mol/cm^3$, $r_3$ is the $H^+$ ion production rate in $mol/(cell*s)$, $d_3$ is the $H^+$ ion uptake rate in $s^{-1}$, $H_0$ is the $H^+$ ion concentration in serum, and $D_3$ is the $H^+$ ion diffusion constant in $cm^2/s$.

Equations (4) and (5) can be rewritten in terms of the variables

$$\eta_1 = N_1 / K_1,$$

$$\eta_2 = N_2 / K_2,$$

$$\Lambda = H / H_0,$$

$$\tau = r_1 t,$$

$$\xi = \sqrt{r_1 / D_N} x. \tag{6}$$

In dimensionless form Eqs. (4) and (5) become

$$\frac{\partial \eta_1}{\partial \tau} = \eta_1 (1 - \eta_1 - \eta_2) - \phi_1(\Lambda) \eta_1 + \nabla_\xi \cdot [(1 - \eta_1 - \eta_2) \nabla_\xi \eta_1],$$

$$\frac{\partial \eta_2}{\partial \tau} = \rho_2 \eta_1 (1 - \eta_1 - \eta_2) - \phi_2(\Lambda) \eta_2 + \nabla_\xi \cdot [(1 - \eta_1 - \eta_2) \nabla_\xi \eta_2],$$

$$\frac{\partial \Lambda}{\partial \tau} = \rho_3 \eta_2 - \delta_3 (\Lambda - 1) + \Delta_3 \nabla_\xi^2 \Lambda, \tag{7}$$

where $\rho_2 = r_2 / r_1$, $\rho_3 = r_3 K_2 / (H_0 r_1)$, $\delta_3 = d_3 / r_1$, and $\Delta_3 = D_3 / D_N$. The death-rate functions are dimensionless and have the form

$$\phi_{1,2}(\Lambda) = \delta_{1,2}[1 - \exp\{-(\frac{\Lambda - \Lambda_{1,2}^{opt}}{2\Lambda_{1,2}^{width}})^2\}], \tag{8}$$

where $\delta_{1,2} = d_{1,2}/r_1$, $\Lambda_{1,2}^{opt} = H_{1,2}^{opt}/H_0$, and $\Lambda_{1,2}^{width} = H_{1,2}^{width}/H_0$ are all dimensionless as well.

For *in vitro* spheroids, doubling times are between 1 and 4 days. Therefore, this yields $r_2 = \ln 2/2.5 \ days \approx 3.2 \times 10^{-6}/s$. For normal tissue wound healing, 4 days seem reasonable for the doubling time; thus, $r_1 = \ln 2/4.0 \ days \approx 2.0 \times 10^{-6}/s$. The volume-limited carrying capacities of tumor and normal tissues are assumed to be the same and given by $K_1 = K_2 \approx 5 \times 10^8 \ cells/cm^3$. For vascular evacuation without buffering $d_3 = \alpha p$, where $\alpha \approx 200/cm$ is the vessel areal density and $p \approx 1.2 \times 10^{-4} \ cm/s$ is the vessel permeability for lactate, resulting in a removal rate of $2.4 \times 10^{-2}/s$. Local buffering might increase this by 25%. Thus, the final estimate for this rate is $d_3 \approx 3.0 \times 10^{-2}/s$. If the serum $pH_0 = 7.4$ is also the optimal $pH$ for normal tissue growth, then $H_1^{opt} = H_0 = 3.98 \times 10^{-11} \ mol/cm^3$. An optimal $pH$ of 6.8 for tumor growth gives $H_2^{opt} = 1.58 \times 10^{-10} \ mol/cm^3$.

As for acid production rate, if a tumor is sufficiently large, then the temporal and spatial derivatives at its core are small. From Eq. (5) we see that $r_3 \approx d_3(H_{core} - H_0)/K_2$. Assuming a core $pH$ of 6.4, this yields $r_3 \approx 2.2 \times 10^{-20} \ mol/(cell*s)$. This last value is remarkably consistent with the curve fit results of the Martin-Jain data to the old model [50]. The lactic acid and cellular diffusion constants are $D_3 \approx 5 \times 10^{-6} \ cm^2/s$ and $D_N \approx 2 \times 10^{-10} \ cm^2/s$, respectively. Using the values above, the numerical estimates of the dimensionless parameters in Eqs. (7) and (8) are as follows: $\rho_2 = r_2/r_1 = 1.6$, $\rho_3 = r_3 K_2/H_0 r_1 = 1.4 \times 10^3$, $\delta_3 = d_3/r_1 = 1.5 \times 10^4$, $\Delta_3 = D_3/D_N = 2.5 \times 10^4$, $\Lambda_1^{opt} = H_1^{opt}/H_0 = 1.0$, $\Lambda_2^{opt} = H_2^{opt}/H_0 = 4.0$, $\Lambda_1^{width} = H_1^{width}/H_0 = 0.1$, $\Lambda_2^{width} = H_2^{width}/H_0 = 0.4$, $\delta_1 = d_1/r_1 = 2.0$, and $\delta_2 = d_2/r_1 = 2.0$. The last four values are admittedly guesswork; but it is reasonable to choose the saturated death rates greater than the growth rates and the width of the optimal $pH$ region for tumor cells greater than that for normal cells.

## 4. Equilibrium densities and dynamical stability analysis

Here, we present a detailed technique we developed to solve for the fixed points of the reaction-diffusion system defined in Eqs. (7), with the sample numerical values of the biological input parameters given in Sec. 3, and determine their dynamical stability. The significance of the fixed points is that they determine equilibrium densities of the normal, tumor, and acid populations at which they could coexist in three-dimensional space (if no malignant behavior is already in progress). The dynamical stability of the fixed points can be determined by solving numerically for the eigenvalues of the Jacobian matrix. If all eigenvalues are negative, the fixed point is considered to be stable; however, if one or more of the eigenvalues is (are) positive, the fixed point is considered to be unstable. A stable fixed point represents an intermediate balance between the normal, tumor, and acid populations where none of them is likely to undergo any noticeable growth into the other two; however, an unstable fixed point where tumor cells are very likely to invade normal tissue upon any slight perturbations of the cellular densities could potentially develop into a malignancy. Note that in Eqs. (7), we present multiple reaction and diffusion factors that could contribute to a change in the cellular densities of the three populations. By mathematical definition, if the fixed point is stable, after slight changes in the equilibrium cellular densities (due to e.g. cellular diffusion) have taken place, the cellular densities tend to converge infinitesimally back to their initial equilibrium values (so the time rates of change of the cellular densities remain at zero after all); however, on the other hand, in an unstable fixed point, after the occurrence of slight perturbations in the equilibrium cellular densities, significant divergences off the equilibrium values are very likely to take place (that is, the cellular populations are not necessarily expected to converge back to their initial values), where malignancies are very likely to develop upon any small perturbations of unstable fixed points; because the presence of the acid specifically would always significantly catalyze the process in the direction where tumor invasion into normal tissue takes place (not vice versa). Note that, when the microenvironment is extremely acidic, the death rate of tumor cells saturates at a value considerably lower than its counterpart for normal cells, and this is interpreted as tumor cells being more capable of surviving in an extremely acidic microenvironment [2,4].

We find the fixed points and analyze their dynamical stability as follows. If we restrict our attention to solutions $\eta_1$, $\eta_2$, and $\eta_3 \equiv \Lambda$, which depend on time but are independent of the spatial coordinate $\xi$, Eqs. (7) reduce to three coupled ordinary differential equations

$$\frac{d\eta_1}{dt} = f_1(\eta_1, \eta_2, \eta_3),$$

$$\frac{d\eta_2}{dt} = f_2(\eta_1, \eta_2, \eta_3),$$

$$\frac{d\eta_3}{dt} = f_3(\eta_1, \eta_2, \eta_3). \tag{9}$$

The fixed points or points of equilibrium are points $(\eta_1^*, \eta_2^*, \eta_3^*)$ where the right sides of Eqs. (9) vanish. Near a fixed point, the differential equations may be linearized in the form

$$\frac{d(\eta_i - \eta_i^*)}{dt} = \sum_{j=1}^{3} \frac{\partial f_i(\eta_1^*, \eta_2^*, \eta_3^*)}{\partial \eta_j}(\eta_j - \eta_j^*). \tag{10}$$

Substituting $\eta_j - \eta_j^* = a_j e^{\lambda t}$, one finds $n$ exponential solutions, with $\lambda$ and $a_j$ given by the eigenvalues and eigenvectors of the matrix $\partial f_i(\eta_1^*,...,\eta_N^*)/\partial n_j$. If all $n$ of the eigenvalues $\lambda$ are negative, $\eta_j - \eta_j^*$ decays exponentially with time, and the fixed point is "stable". If one or more of the $\lambda$ values is (are) positive on the other hand, the fixed point is unstable. Note that variables $\eta_1(t)$, $\eta_2(t)$, and $\Lambda(t)$ depend on $t$ but not on the coordinate $\xi$.

Next, we solve numerically for the fixed points. To find the fixed points, we set the derivatives in the cellular equations [the first and second of Eqs. (7)] equal to zero after eliminating $\Lambda$ using the acid equation [the third in Eqs. (7)]. This yields two non-linear equations in two unknowns, namely $\eta_1$ and $\eta_2$, which are physically contained within the interval [0,1]. Therefore, we partition the domain $0 \leq \eta_1 \leq 1 \otimes 0 \leq \eta_2 \leq 1$ into a rectangular grid with $\Delta\eta_1 = \Delta\eta_2 = 0.01$ (10000-points grid), and use these 10000 points as numerical inputs for a multi-dimensional Newton's algorithm. For a specific Newton's solution for $\eta_1$ and $\eta_2$, $\eta_3$ is found using the third equation in (7). We determine the stability of the resulting solutions by numerically computing the eigenvalues of the Jacobian matrix (we have published the software package we used to perform the dynamical stability analysis presented here [4]). The fixed-points analysis for the parameters given in the previous section is shown below [4,51]:

| Fixed Point | Eigenvalues | Stable/Unstable |
|---|---|---|
| (− 0.26500634, 0.01500634, 1.14005922) | (− 1.500000e+04, 1.205189e−01, 1.205189e−01) | Unstable (unphysical) |
| (− 0.23499366, − 0.01500634, 0.85994078) | (− 1.500000e+04, 1.295203e−01, 1.295203e−01) | Unstable (unphysical) |
| (0.00000000, 0.37206032, 4.47256298) | (− 1.498978e+04, − 1.081070e+01, − 1.372060e+00) | Stable |
| (0.00000000, 0.26431209, 3.46691282) | (− 1.500676e+04, 6.337857e+00, − 1.264312e+00) | Unstable |
| (0.00000000, − 0.25000000, − 1.33333333) | (− 1.500000e+04, − 7.500000e−01, 4.000000e−01) | Unstable (unphysical) |

**Table 1.** Fixed-point analysis for the biological input parameters given in Sec. 3.

Fixed point 3 is stable (all eigenvalues are negative). Fixed point 4 is unstable (some eigenvalues are positive) and is very likely to develop into a malignancy upon any slight perturbations of the equilibrium cellular densities [note that when $\eta_1$ is zero, this means that all the normal population has died at the interface (microenvironment) joining the three populations]. Fixed points 1, 2, and 5 are all unstable and unphysical (negative density values). Note that the existence of two different physical solutions (either stable or unstable) means that for the biological input parameters given in Sec. 3, wherever the normal, tumor, and acid populations are found in a dynamical equilibrium in three-dimensional space, it must be in one of these two solutions [note that we are not reporting here trivial fixed points such as (0,0,1) and (1,0,1)].

## 5. Discussion, conclusions, and future work

The canonical Gatenby-Gawlinski model of the acid-mediated tumor invasion hypothesis presented in Sec. 3 shows how the densities of normal cells, tumor cells, and excess $H^+$ ions change due to two factors: first, local chemical reactions between the three populations and, second, density-dependent diffusion by which these populations spread out in space. According to the hypothesis, the dominant mechanism for cancer invasion is the tumor-induced alteration of microenvironmental $pH$. Moreover, the hypothesis predicts an acidic $pH$ gradient extending into the peritumoral normal tissue, where the normal mammalian cells (but not tumor cells) are intolerant of acidic interstitial $pH$ in the range typically found within this gradient.

In Sec. 4 we have developed detailed analytical and computational techniques to analyze the dynamical stability of the Gatenby-Gawlinski reaction-diffusion system by solving for the fixed points of the differential equations [see Eqs. (7)] that represent how the normal, tumor, and acid populations evolve with time, then determining the stability of these fixed points by solving for the eigenvalues of the Jacobian matrix. The significance of the fixed points is that they determine possible densities at which the normal, tumor, and acid populations could coexist in dynamical equilibrium in three-dimensional space (if no malignant behavior is already in progress; that is, the time rates of change of the densities of the three populations are equal to zero). As for the dynamical stability of the fixed points, if all eigenvalues are negative, the fixed point is stable; however, if one or more of the eigenvalues is (are) positive, the fixed point is unstable. In a stable fixed point, after slight changes in the equilibrium cellular densities have taken place (due to e.g. cellular diffusion), the cellular densities tend to converge infinitesimally back to their initial equilibrium values (so the time rates of change of the cellular densities remain at zero after all); however, on the other hand, in an unstable fixed point, after the occurrence of slight changes in the equilibrium densities, significant divergences off the initial equilibrium values are likely to take place, where malignancies are likely to emerge; because, on balance, the presence of the acid specifically would always significantly corroborate the process into only one direction; that is, tumor invasion into normal tissue (not vice versa). Whenever a malignant behavior is already in progress, this implies that the time rate of change of the density of tumor cells is positive. A fixed point (either stable or unstable) physically represents an intermediate balance between the three populations where none of them is undergoing any noticeable growth into the other two. If the fixed point is stable, a malignant behavior is unlikely to emerge, even after the occurrence of slight perturbations of the equilibrium cellular densities at the interface (microenvironment) joining the three populations. On the other hand, an unstable fixed point where tumor cells are very likely to invade normal tissue upon any slight perturbations of the equilibrium cellular densities could potentially develop into a malignancy. Note that, in the reaction-diffusion model we present here, from a mathematical standpoint and within the acid-mediation hypothesis, only the dynamical stability of the equilibrium points (not the initial equilibrium density values) controls whether a prospective malignant behavior is likely; because the dynamical stability by itself controls what happens after any slight perturbations of the initial equilibrium density values have taken place. We have solved numerically for the fixed points (and their dynamical stability) of the reaction-diffusion system presented in Eqs. (7) with the sample numerical values of the biological input parameters provided in Sec. 3. As demonstrated in Table 1, for physical fixed points (all density values are positive), our numerical results show that both stable (fixed point 3) and unstable (fixed point 4) configurations are possible. For the particular set of the biological input parameters we have operated on in Sec. 3, the existence of two different physical solutions means that wherever the three populations are found in a dynamical equilibrium in three-dimensional space, it must be in one of these two solutions.

As for future work, it would be interesting to study systematically how the fixed-point behavior depends on the biological input parameters. In simple systems this can be done analytically; but in systems as complex as the one introduced in Eqs. (7), numerical methods are needed. Possibly, one can use numerical algorithms [52], designed to compute branches of stable and unstable solutions, and compute the Floquet multipliers that determine stability along these branches. Starting data for the computation of periodic orbits are generated automatically at the Hopf bifurcation points. Such techniques can also locate folds, branch points, period-doubling bifurcations, and bifurcations to tori. Along branches of periodic solutions, branch switching is possible at branch points and at period-doubling bifurcations. Once the bifurcation diagrams in parameter space are understood in this way, dynamical simulations can be performed to determine tumor growth or regression rates.

## Acknowledgment

I am grateful to Dr. Edward Gawlinski and Dr. Theodore Burkhardt for very thoughtful comments and discussions.

## References

[1] Gatenby, R.A., 1991. Population ecology issues in tumor growth. *Cancer Research*, *51*(10), pp.2542-2547.

[2] Gatenby, R.A. and Gawlinski, E.T., 1996. A reaction-diffusion model of cancer invasion. *Cancer research*, *56*(24), pp.5745-5753.

[3] Patel, A.A., Gawlinski, E.T., Lemieux, S.K. and Gatenby, R.A., 2001. A cellular automaton model of early tumor growth and invasion: the effects of native tissue vascularity and increased anaerobic tumor metabolism. *Journal of Theoretical Biology*, *213*(3), pp.315-331.

[4] Fouad, A., 2015. *Density-dependent diffusion models with biological applications: Numerical modeling and analytic analysis* (Doctoral dissertation, TEMPLE UNIVERSITY).

[5] Warburg O. *The metabolism of tumors*. London: Constable Press, 1930.

[6] Di Chiro, G., Hatazawa, J., Katz, D.A., Rizzoli, H.V. and De Michele, D.J., 1987. Glucose utilization by intracranial meningiomas as an index of tumor aggressivity and probability of recurrence: a PET study. *Radiology*, *164*(2), pp.521-526.

[7] Haberkorn, U., Strauss, L.G., Reisser, C.H., Haag, D., Dimitrakopoulou, A., Ziegler, S., Oberdorfer, F., Rudat, V. and Van Kaick, G., 1991. Glucose uptake, perfusion, and cell proliferation in head and neck tumors: relation of positron emission tomography to flow cytometry. *J Nucl Med*, *32*(8), pp.1548-55.

[8] Hawkins, R.A., Hoh, C., Glaspy, J., Choi, Y., Dahlbom, M., Rege, S., Messa, C., Nietszche, E., Hoffman, E., Seeger, L. and Maddahi, J., 1992, October. The role of positron emission tomography in oncology and other whole-body applications. In *Seminars in nuclear medicine* (Vol. 22, No. 4, pp. 268-284). WB Saunders.

[9] Patz Jr, E.F., Lowe, V.J., Hoffman, J.M., Paine, S.S., Harris, L.K. and Goodman, P.C., 1994. Persistent or recurrent bronchogenic carcinoma: detection with PET and 2-[F-18]-2-deoxy-D-glucose. *Radiology*, *191*(2), pp.379-382.

[10] Yonekura Y, Benua RS, Brill AB, Som P, Yeh SD, Kemeny NE, Fowler JS, MacGregor RR, Stamm R, Christman DR, Wolf AP. Increased accumulation of 2-deoxy-2[18F]fluoro-D-glucose in liver metastasis from colon cancer. *Journal of Nuclear Medicine*, 23, 1982, pp. 1133-1137.

[11] Griffiths, J.R., 1991. Are cancer cells acidic?. *British journal of cancer*, *64*(3), p.425.

[12] Gillies, R.J., Liu, Z. and Bhujwalla, Z., 1994. 31P-MRS measurements of extracellular pH of tumors using 3-aminopropylphosphonate. *American Journal of Physiology-Cell Physiology*, *267*(1), pp.C195-C203.

[13] Martin, G.R. and Jain, R.K., 1994. Noninvasive measurement of interstitial pH profiles in normal and neoplastic tissue using fluorescence ratio imaging microscopy. *Cancer research*, *54*(21), pp.5670-5674.

[14] Martin GR. *Thesis*; Pittsburgh, PA: Carnegie Mellon University, Pittsburgh, 1995.

[15] Stubbs, M., Rodrigues, L., Howe, F.A., Wang, J., Jeong, K.S., Veech, R.L. and Griffiths, J.R., 1994. Metabolic consequences of a reversed pH gradient in rat tumors. *Cancer research*, *54*(15), pp.4011-4016.

[16] Park, H.J., Lyons, J.C., Ohtsubo, T. and Song, C.W., 1999. Acidic environment causes apoptosis by increasing caspase activity. *British journal of cancer*, *80*(12), p.1892.

[17] Williams, A.C., Collard, T.J. and Paraskeva, C., 1999. An acidic environment leads to p53 dependent induction of apoptosis in human adenoma and carinoma cell lines: Implications for clonal selection during colorectal carcinogenesis. *Oncogene*, *18*(21), pp.3199-3204.

[18] Rubin, H., 1971. pH and population density in the regulation of animal cell multiplication. *The Journal of cell biology*, *51*(3), pp.686-702.

[19] Dairkee, S.H., Deng, G., Stampfer, M.R., Waldman, F.M. and Smith, H.S., 1995. Selective cell culture of primary breast carcinoma. *Cancer research*, *55*(12), pp.2516-2519.

[20] Casciari, J.J., Sotirchos, S.V. and Sutherland, R.M., 1992. Variations in tumor cell growth rates and metabolism with oxygen concentration, glucose concentration, and extracellular pH. *Journal of cellular physiology*, *151*(2), pp.386-394.

[21] Bierman, A.J., Tertoolen, L.G., De Laat, S.W. and Moolenaar, W.H., 1987. The Na+/H+ exchanger is constitutively activated in P19 embryonal carcinoma cells, but not in a differentiated derivative.

Responsiveness to growth factors and other stimuli. *Journal of Biological Chemistry*, *262*(20), pp.9621-9628.

[22] Kaplan, D.L. and Boron, W.F., 1994. Long-term expression of cH-ras stimulates Na-H and Na (+)-dependent Cl-HCO3 exchange in NIH-3T3 fibroblasts. *Journal of Biological Chemistry*, *269*(6), pp.4116-4124.

[23] Vaupel, P., Kallinowski, F. and Okunieff, P., 1989. Blood flow, oxygen and nutrient supply, and metabolic microenvironment of human tumors: a review. *Cancer research*, *49*(23), pp.6449-6465.

[24] Smith, T.A.D., 2000. Mammalian hexokinases and their abnormal expression in cancer. *British journal of biomedical science*, *57*(2), p.170.

[25] Kato, Y., Nakayama, Y., Umeda, M. and Miyazaki, K., 1992. Induction of 103-kDa gelatinase/type IV collagenase by acidic culture conditions in mouse metastatic melanoma cell lines. *Journal of Biological Chemistry*, *267*(16), pp.11424-11430.

[26] Rozhin, J., Sameni, M., Ziegler, G. and Sloane, B.F., 1994. Pericellular pH affects distribution and secretion of cathepsin B in malignant cells. *Cancer research*, *54*(24), pp.6517-6525.

[27] Yamasaki, H., 1991. Aberrant expression and function of gap junctions during carcinogenesis. *Environmental health perspectives*, *93*, p.191.

[28] Walenta, S., Wetterling, M., Lehrke, M., Schwickert, G., Sundfør, K., Rofstad, E.K. and Mueller-Klieser, W., 2000. High lactate levels predict likelihood of metastases, tumor recurrence, and restricted patient survival in human cervical cancers. *Cancer research*, *60*(4), pp.916-921.

[29] Wolfram, S., 1986. *Theory and applications of cellular automata* (Vol. 1, pp. x+-560). Singapore: World scientific.

[30] Düchting, W. and Vogelsaenger, T., 1984. Analysis, forecasting, and control of three-dimensional tumor growth and treatment. *Journal of medical systems*, *8*(5), pp.461-475.

[31] Düchting, W. and Vogelsaenger, T., 1985. Recent progress in modelling and simulation of three-dimensional tumor growth and treatment. *Biosystems*, *18*(1), pp.79-91.

[32] Qi, A.S., Zheng, X., Du, C.Y. and An, B.S., 1993. A cellular automaton model of cancerous growth. *Journal of theoretical biology*, *161*(1), pp.1-12.

[33] Smolle, J. and Stettner, H., 1993. Computer simulation of tumour cell invasion by a stochastic growth model. *Journal of theoretical biology*, *160*(1), pp.63-72.

[34] Smolle, J., 1998. Fractal tumor stromal border in a nonequilibrium growth model. *Analytical and quantitative cytology and histology/the International Academy of Cytology [and] American Society of Cytology*, *20*(1), pp.7-13.

[35] Smolle, J., 1998. Cellular automaton simulation of tumour growth–equivocal relationships between simulation parameters and morphologic pattern features. *Analytical Cellular Pathology*, *17*(2), pp.71-82.

[36] Kansal, A.R., Torquato, S., Harsh, G.R., Chiocca, E.A. and Deisboeck, T.S., 2000. Simulated brain tumor growth dynamics using a three-dimensional cellular automaton. *Journal of theoretical biology*, *203*(4), pp.367-382.

[37] Kansal, A.R., Torquato, S., Harsh, G.R., Chiocca, E.A. and Deisboeck, T.S., 2000. Simulated brain tumor growth dynamics using a three-dimensional cellular automaton. *Journal of theoretical biology*, *203*(4), pp.367-382.

[38] Chaplain, M.A. and Anderson, A.R., 1995. Mathematical modelling, simulation and prediction of tumour-induced angiogenesis. *Invasion & metastasis*, *16*(4-5), pp.222-234.

[39] Anderson, A.R. and Chaplain, M.A.J., 1998. Continuous and discrete mathematical models of tumor-induced angiogenesis. *Bulletin of mathematical biology*, *60*(5), pp.857-899.

[40] Markus, M., Böhm, D. and Schmick, M., 1999. Simulation of vessel morphogenesis using cellular automata. *Mathematical biosciences*, *156*(1), pp.191-206.

[41] Adam, J.A., 1997. General aspects of modeling tumor growth and immune response. In *A survey of models for tumor-immune system dynamics* (pp. 15-87). Birkhäuser Boston.

[42] Olsen, L., Sherratt, J.A., Maini, P.K. and Arnold, F., 1997. A mathematical model for the capillary endothelial cell-extracellular matrix interactions in wound-healing angiogenesis. *Mathematical Medicine and Biology*, *14*(4), pp.261-281.

[43] Ward, J.P. and King, J.R., 1997. Mathematical modelling of avascular-tumour growth. *Mathematical Medicine and Biology*, *14*(1), pp.39-69.

[44] Ward, J.P. and King, J.R., 1999. Mathematical modelling of avascular-tumour growth II: modelling growth saturation. *Mathematical Medicine and Biology*, *16*(2), pp.171-211.

[45] Perumpanani, A.J. and Norbury, J., 1999. Numerical interactions of random and directed motility during cancer invasion. *Mathematical and computer modelling*, *30*(7), pp.123-133.

[46] Sleeman, B.D., Anderson, A.R.A. and Chaplain, M.A.J., 1999. A mathematical analysis of a model for capillary network formation in the absence of endothelial cell proliferation. *Applied mathematics letters*, *12*(8), pp.121-127.

[47] Jackson, T.L. and Byrne, H.M., 2000. A mathematical model to study the effects of drug resistance and vasculature on the response of solid tumors to chemotherapy. *Mathematical biosciences*, *164*(1), pp.17-38.

[48] Byrne, H.M. and Chaplain, M.A.J., 1995. Growth of nonnecrotic tumors in the presence and absence of inhibitors. *Mathematical biosciences*, *130*(2), pp.151-181.

[49] Pao, C.V., 2014. A Lotka–Volterra cooperating reaction–diffusion system with degenerate density-dependent diffusion. *Nonlinear Analysis: Theory, Methods & Applications*, *95*, pp.460-467.

[50] Martin, G.R. and Jain, R.K., 1994. Noninvasive measurement of interstitial pH profiles in normal and neoplastic tissue using fluorescence ratio imaging microscopy. *Cancer research*, *54*(21), pp.5670-5674.

[51] Fouad, A.M., 2018. A kinetic view of acid-mediated tumor invasion. *European Biophysics Journal*, *47*(2), pp.185-189.

[52] Doedel, E.J., Paffenroth, R.C., Champneys, A.R., Fairgrieve, T.F., Kuznetsov, Y.A., Oldeman, B.E., Sandstede, B. and Wang, X., 2002. AUTO 2000: CONTINUATION AND BIFURCATION SOFTWARE.